\documentclass[12pt, A4]{article}
\usepackage[utf8]{inputenc}
\usepackage[a4paper]{geometry}
\usepackage{graphicx}
\usepackage{amsmath}
\usepackage{natbib}
\usepackage{graphicx}
\usepackage{amsmath}
\usepackage{txfonts}
\usepackage{array}
\usepackage{float}
\usepackage{bbold}
\usepackage{natbib}
\usepackage{comment}
\usepackage{authblk}
\usepackage[normalem]{ulem}
\newtheorem{theorem}{Theorem}

\title{Causal inference with a functional outcome}

\author{Kreske Ecker\footnote{Corresponding author: kreske.ecker@umu.se}\ }
\author{Xavier de Luna}
\author{Lina Schelin}

\affil{Department of Statistics, \\ Umeå School of Business, Economics and Statistics, \\ Umeå University, Umeå, Sweden}

\date{}

\begin{document}
\maketitle

\begin{abstract}
	
This paper presents methods to study the causal effect of a binary treatment on a functional outcome with observational data. We define a Functional Average Treatment Effect and develop an outcome regression estimator. We show how to obtain valid inference on the FATE using simultaneous confidence bands, which cover the FATE with a given probability over the entire domain. Simulation experiments illustrate how the simultaneous confidence bands take the multiple comparison problem into account. Finally, we use the methods to infer the effect of early adult location on subsequent income development for one Swedish birth cohort.

\end{abstract}
\section{Introduction}
\label{sec:introduction}
The impact of economic conditions at labour market entry on subsequent wage development and other labour market outcomes has been the focus of a range of recent studies; see e.g., \cite{Altonji, Kwon, Oreo2, Raaum, Schwandt, Aslund}. In this paper, we contribute by utilising large-scale register data providing much longer follow-up periods, and by developing methods that allow us to draw valid inference in this context, i.e. taking into account an inherent multiple comparison problem regarding the causal effect of initial labour market conditions on income observed through whole working lives. 

\begin{figure}[h!]
\includegraphics[width=0.9\linewidth]{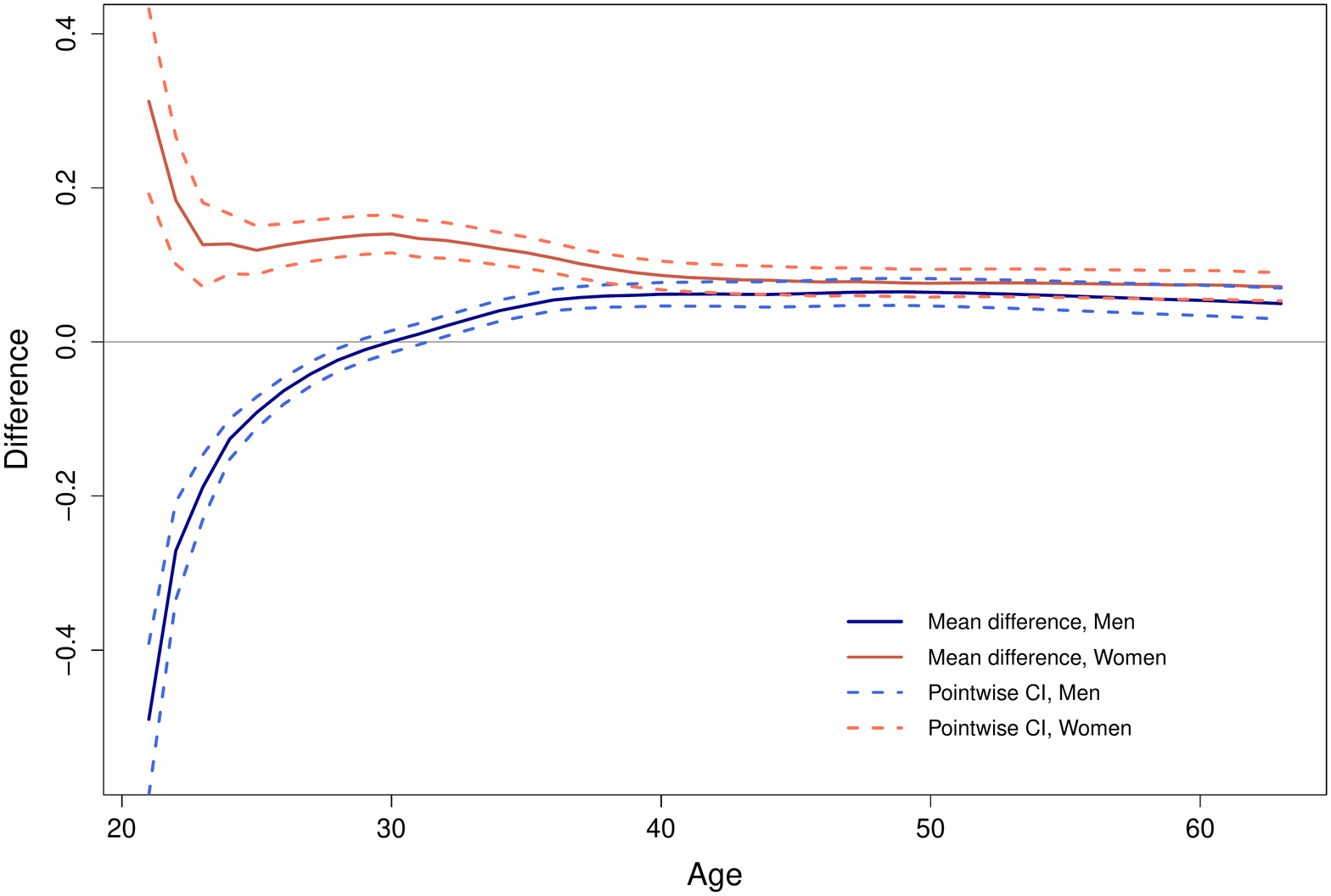} 
\caption{Mean differences in LCI (urban versus rural location at age 20), together with 95\% pointwise confidence bands, for the 1954 birth cohort in Sweden. By gender, blue: men, red: women.}
\label{fig:Mean incomes}
\end{figure}

More precisely, we investigate the causal effect of early adult residence location on subsequent cumulative lifetime incomes for the 1954 Swedish birth cohort, based on population-wide register data. The data available allows us to follow income development through most of the cohort's working life: from adolescence until the age of 63. We define the binary treatment of interest as living in either an urban or a rural area of Sweden at the age of 20 (in 1974). The outcome, logarithmized cumulative income (LCI), is observed annually. Figure \ref{fig:Mean incomes} shows the mean differences in LCI between the two treatment groups (urban - rural location at age 20), computed separately for men and women, over the period 1975-2017, together with 95\% pointwise confidence bands. This comparison is naive in two aspects: first, it has no causal interpretation since confounding variables are not controlled for; and second, the pointwise confidence bands are not meaningful because of the multiple comparison problem, i.e. the displayed bands yield only pointwise control of the coverage probability, whereas the actual nominal coverage over the entire domain of the income functions is considerably lower.

The observed incomes are assumed to arise from an underlying continuous income accumulation process, from which we take discrete observations. This motivates the use of methods from the field of functional data analysis (FDA) \citep[see, e.g.,][for an overview]{Ramsay, Wang}. While FDA methods are not the only way to analyse such data, they have a major advantage in that they allow to address the multiple comparison problem in an elegant and practical fashion as described below. 
To deal with a non-randomised treatment assignment (here early adult residence location), we use the potential outcome framework \citep{RoRu,holland86} to define a causal parameter, the Functional Average Treatment Effect (FATE), which takes the functional nature of the outcome into account. Up to our knowledge, there is only one earlier attempt \citep{Belloni} to formalise and study causal inference in the context of a functional outcome. Our methodological contribution differs in that we provide exact finite sample inference, while \citet{Belloni} provide justification for asymptotic bootstrap inference.
Other related works include \citep{Chib, Jacobi} in which the terminology "panel outcomes" is used. In this literature, the multiple comparison problem is not dealt with, although it is also less severe since typically only a few time points of follow up are considered. Note that complex causal effects can also be defined when scalar potential outcomes are compared, e.g., by contrasting their distribution functions \citep{Linetal:2022}.

 We propose and study an estimator of the FATE based on outcome regression \citep{Tan}. Quantifying the uncertainty in this estimation presents a challenge since existing inferential techniques developed for univariate outcomes cannot satisfactorily address the multiple comparison problem induced by the functional nature of the causal parameter in this setting. Since the functional parameter is infinite-dimensional, standard techniques to control for multiple testing  \citep[e.g.][]{Holm} would be too conservative. Instead, techniques within the field of FDA provide a variety of solutions to the multiple comparisons problem, including, e.g., approaches for p-value adjustments in hypothesis testing \citep{vsevolozhskaya2014pairwise, Pini17, Abram} as well as simultaneous confidence bands based on results from random field theory \citep{Liebl, Liebl2, Telschow}. The former approaches rely on permutation tests rather than distributional assumptions and are therefore computationally intensive. We here expand the methods and results obtained by \cite{Liebl, Liebl2} to deal with a functional causal parameter estimated with an outcome regression estimator, and thereby obtain simultaneous confidence bands for the FATE. These simultaneous confidence bands control the probability that the true FATE lies outside the bands over its entire domain and thus provide global control of the coverage probability. In addition, they are less computationally intensive than many simulation-based approaches, and their widths can adapt to the local structure of the data \citep{Liebl}.  

The rest of the paper is organised as follows. Section 2 presents the theory and methods for drawing exact finite sample inference using simultaneous confidence bands for the functional causal parameter of interest.
Section 3 presents the study of the effect of early adult residence location on cumulative lifetime incomes using the methods introduced. Numerical experiments inspired from the real data application are used in Section 4 to evaluate the performance of the simultaneous confidence bands when the covariance function is estimated. Section 5 concludes the paper with a discussion.

\section{Theory: Functional average treatment effect}

\subsection{Definition and identification}\label{identification.sec}
As mentioned above, our goal is to estimate the causal effect of early adult residence location on income accumulation over time for one Swedish birth cohort. This example will be used throughout the section for illustration. 

To simplify notation, we do not distinguish between random variables and their realisations.  Let $z_i$ be a binary treatment indicator for individual $i$, taking the value $z_i=1$ if individual $i$ is exposed to treatment, and $z_i=0$ otherwise.
In the motivating example, this would correspond to living in an urban and a rural location, respectively, at the age of 20.
Corresponding to these treatment levels, we define the two potential outcome processes $y_{1i}=\{y_{1i}(t):t \in [0,1]\}$ and $y_{0i}=\{y_{0i}(t):t \in [0,1]\}$, here cumulative incomes over time. These processes are assumed to be once continuously differentiable almost surely and without loss of generality, the domain of the argument $t$ has been scaled to $[0,1]$. 
For each individual, only one of these potential outcome processes can be observed, depending on whether they received the treatment or not. The actually observed outcome process is assumed to be $y_i(t) = y_{1i}(t)\,z_i + y_{0i}(t)\,(1-z_i).$ This, together with the fact that treatment assignment for one unit does not affect other units' potential outcomes is often called the "stable unit treatment value assumption (SUTVA)" \citep[][Chap 1]{imbensrubinbook}. In our application, this entails that individuals' choice of residence location at age 20 does not affect other's lifetime income processes. While violation of such an assumption cannot be ruled out empirically, it is reasonable to assume that violation of SUTVA, i.e. interference between individuals, are negligible since the regions considered are large in area and population, and we focus only on one cohort.
In addition, let $\boldsymbol{x}_i= (1, x_{1i}, \dots, x_{Ki})^T$ be a vector of covariates measured prior to treatment assignment, i.e., they are assumed not to be affected by treatment. This vector includes a 1-element for ease of notation later on. We have a random sample of size $n$, with $i=1,\ldots,n$, for which we observe $\boldsymbol{x}_i, z_i$ and $y_i = \{y_i(t): t \in [0,1]\}$. Let $n_1$ and $n_0= n-n_1$ be the number of treated and untreated units in this sample, respectively. 

The focus in this paper is on one causal parameter, the Functional Average Treatment Effect (FATE):
\begin{equation*}
  \theta := \left\{\theta(t) =\mathbb{E}\left( y_{1i}(t)-y_{0i}(t)\right):t \in [0,1]\right\}.  
\end{equation*}
An alternative causal parameter of interest is the Functional Average Treatment Effect on the Treated population (FATT). The developments for this parameter are given in Appendix \ref{fatt.sec}.

If the treatment assignment is randomised, the FATE can be identified directly from the observed data. However, this is not the case in our case study or in many other practical applications. Individuals who live in an urban area at the age of 20 could differ from those living in a rural area with respect to characteristics which in turn relate to income accumulation. 
In the presence of such confounding, identification of the causal effect from the observed data requires further assumptions \citep[e.g.][]{wooldridgebook,holland86,RoRu}. Here we use the following identification assumptions:

\noindent\textsc{Assumption 1 (Overlap)}
\begin{align*}
   \mbox{a.}\ \ \ \ \ &\Pr(z_i = 0\mid \boldsymbol{x}_i)>0, \ \ \ \ \ \forall \boldsymbol{x}_i, \\  
   \mbox{b.}\ \ \ \ \ &\Pr(z_i = 1\mid \boldsymbol{x}_i)>0, \ \ \ \ \ \forall \boldsymbol{x}_i.
\end{align*}
\

\noindent\textsc{Assumption 2 (Mean ignorability)}
\begin{align*}
    \mbox{a.}\ \ \ \ \ &\mathbb{E}\left(y_{0i}(t) \mid z_i, \boldsymbol{x}_i \right) = \mathbb{E}\left(y_{0i}(t) \mid \boldsymbol{x}_i \right) \ \ \ \ \ \forall t\in[0,1], \\
    \mbox{b.}\ \ \ \ \ &\mathbb{E}\left(y_{1i}(t) \mid z_i, \boldsymbol{x}_i \right) = \mathbb{E}\left(y_{1i}(t) \mid \boldsymbol{x}_i \right) \ \ \ \ \ \forall t\in[0,1]. 
\end{align*}
Mean ignorability \citep[][Sec. 21.3]{wooldridgebook} is a weaker version of the distribution ignorability assumption often used in the literature, where the potential outcomes and the treatment assignment are assumed to be independent in distribution conditional on the pre-treatment covariates. When an ignorability assumption holds, we say that $x$ contains all confounding covariates.

The causal parameter $\theta$ is identified under the above assumptions since we can then write, for all $t\in[0,1]$:

\begin{align*}
    \theta(t) &= \mathbb{E}(y_{1i}(t)) -  \mathbb{E}(y_{0i}(t)) \\
    &= \mathbb{E} \big[\mathbb{E}(y_{1i}(t)\,\mid\,z_i = 1, \boldsymbol{x}_i) -  \mathbb{E}(y_{0i}(t)\,\mid\,z_i = 0, \boldsymbol{x}_i) \big] \\
    &= \mathbb{E} \big[\mathbb{E}(y_{i}(t)\,\mid\,z_i = 1, \boldsymbol{x}_i) -  \mathbb{E}(y_{i}(t)\,\mid\,z_i = 0, \boldsymbol{x}_i) \big].
\end{align*}

Assumption 1 implies that there are no combinations of values on the pre-treatment covariates for which an individual would have probability zero of living in either an urban or a rural area at the age of 20. This assumption can be checked empirically as we did with our case study, see Appendix \ref{comparebands} for details. Note that in any case this assumption can be enforced by discarding individuals from the population, changing thereby the population and parameter of interest.
On the other hand, Assumption 2 cannot be empirically corroborated and there is always a possibility that some unobserved characteristic confounds the effect  of individuals' residence location at 20 years on lifetime incomes. For instance, while we will control for educational attainments, such a covariate might not measure all innate skills that could affect both residence location and lifetime incomes. This is a common challenge when drawing causal inferences from observational data, and it means that the results should be interpreted with a certain level of caution.

\subsection{Estimation and inference}\label{estimation.sec}

We consider an outcome-regression estimator for $\theta$, which assumes linear function-on-scalar models \citep{Ramsay} for the outcomes:

\begin{equation}
    y_{zi}(t) = \boldsymbol{x}_i^T \ \boldsymbol{\beta}_z(t) + \varepsilon_{zi}(t), \quad z = 0,1, 
    \label{flm}
\end{equation}
where $\boldsymbol{\beta}_z(t)$ are $K+1$-dimensional functional parameters. Usual diagnostic tools can be used to include covariates that are transformations of the original covariates, although the inference will not then take into account such a model building step.

In order to obtain exact finite sample results below, we make the following assumptions on the random error processes $\varepsilon_{zi} = \{\varepsilon_{zi}(t): t \in [0,1]\}$.

\noindent\textsc{Assumption 3 (Random error processes)}

\noindent The vector $\boldsymbol{\varepsilon}_z=(\varepsilon_{z1},\cdots,\varepsilon_{zn})^T$ of error terms in model \eqref{flm} is a multivariate Gaussian process:
\begin{equation*}
     \boldsymbol{\varepsilon}_z \sim \mathcal{MGP}\left( 0, \sigma_z, \mathbb{I}_{n_z} \right),
\end{equation*}
where $ \sigma_{z}(s,t) = Cov\big(\varepsilon_{zi}(s)\ , \ \varepsilon_{zi}(t)\big), \ s, t \in [0,1]$, and $\mathbb{I}_{n_z}$ is a $n_z \times n_z$ identity matrix. 

Let $\boldsymbol{y}(t) = \big(y_1(t), \dots, y_n(t)\big)^T$ denote the vector of observed outcomes, with $\boldsymbol{y}_1(t) = \{y_i(t) : z_i = 1 \}$ and $\boldsymbol{y}_0(t) = \{y_i(t) : z_i = 0 \}$ its subsets for the treated and untreated units respectively. Similarly, let $\boldsymbol{X}_1$ be the $n_1\times (K+1)$ matrix containing observations on the $K$ covariates (and a column of 1 for the intercept) for the treated subgroup, and $\mathbf{X}_0$ the corresponding $n_0\times (K+1)$ matrix of covariates for the non-treated, i.e., subsets of $\boldsymbol{X} = (\boldsymbol{x}_1, \dots, \boldsymbol{x}_n)^T$. Then, the functional regression coefficients can be estimated by  using ordinary least squares:

\begin{equation}
    \hat{\boldsymbol{\beta}}_z(t) =  (\mathbf{X}^T_{z}\mathbf{X}_z)^{-1}\mathbf{X}_{z}^T\boldsymbol{y}_{z}(t),
    \label{beta}
\end{equation}
$\hat{\boldsymbol{\beta}}_z = \{\hat{\boldsymbol{\beta}}_z(t): t \in [0,1]\}$.

We can then construct an estimator for the FATE by imputing the unobserved potential outcomes with predictions from the fitted outcome models above:
\begin{equation}
    \begin{split}
    \hat{\theta}(t) &  = \frac{1}{n} \sum_{i=1}^n \Big( \boldsymbol{x}_i^T \hat{\boldsymbol{\beta}}_1(t) - \boldsymbol{x}_i^T \hat{\boldsymbol{\beta}}_0(t) \Big) \\
    & =  \overline{\boldsymbol{x}} \hat{\boldsymbol{\beta}}_1(t) - \overline{\boldsymbol{x}} \hat{\boldsymbol{\beta}}_0(t), 
    \label{theta}
    \end{split}
\end{equation}
where $
    \overline{\boldsymbol{x}} = \frac{1}{n}\, \mathbb{1}_{1 \times n} \mathbf{X} 
    $
is the vector of sample means. This yields the estimator  $\hat{\theta}= \{\hat{\theta}(t): t\in [0,1]\}$.

We  now give the finite sample properties of the estimator $\hat{\theta}$ .
Throughout, we assume a fixed design for the regression models in \eqref{flm}, meaning that inference is conditional on $\mathbf{X}$.

\begin{theorem}\label{thm1}
    Let SUTVA, and Assumptions 1-3 hold. 
    Then, the estimator $\hat{\theta}$ is a Gaussian process: 
    \begin{equation}
    \label{dist2}
        \hat{\theta} \mid \mathbf{X}  \sim \mathcal{GP} \left( \overline{\boldsymbol{x}} \boldsymbol{\beta}_1 - \overline{\boldsymbol{x}} \boldsymbol{\beta}_0, C_\theta \right), 
    \end{equation} 
   where
    \begin{equation}
        C_\theta (s,t) = \sigma_1(s,t) \ \overline{\boldsymbol{x}}\left(\mathbf{X}^T_{1} \mathbf{X}_{1}\right)^{-1} \overline{\boldsymbol{x}}^T + \sigma_0(s,t) \ \overline{\boldsymbol{x}}\left(\mathbf{X}^T_{0} \mathbf{X}_{0}\right)^{-1} \overline{\boldsymbol{x}}^T.
    \label{covtheta1}
    \end{equation}
\end{theorem}

See Appendix \ref{GPproof.sec} for a proof. 
Note that the Gaussian condition in Assumption 3 is not needed to derive well known pointwise properties of $\hat{\theta}(t)$ such as consistency for $\theta(t)$ \citep[e.g.,][]{genback2019causal}. We use this assumption to obtain exact finite sample inference on the functional estimator.

Using Theorem \ref{thm1}, we can build on the simultaneous confidence bands presented in \cite{Liebl,Liebl2}, which allow for a global control of the coverage probability over the entire domain, to quantify the uncertainty in the estimation of $\theta$. A $(1-\alpha)$\% simultaneous confidence band for $\theta$ is given by
\begin{equation}
    \hat{\theta}(t) \pm {u}_\alpha(t) \sqrt{C_\theta(t,t)},
    \label{band}
\end{equation}
where 
 $u_\alpha = \{u_\alpha(t):t\in [0,1]\}$ is a continuous threshold function such that, for a given level $\alpha$,

\begin{equation*}
    P \left( \theta(t) \in \hat{\theta}(t) \pm u_\alpha(t) \sqrt{C_\theta(t,t)} \quad \forall \ t \in [0,1] \right) \geq 1 - \alpha.
\end{equation*}

\noindent In other words, (\ref{band}) yields a $(1-\alpha)$ simultaneous confidence band for the functional parameter $\theta$, which is such that $\theta(t)$ lies within the band over the entire domain with a probability of at least $(1-\alpha)$. See \cite{Liebl} for details on how these bands are constructed.
If we were to instead use a naive constant threshold of, e.g., $u_{0.05}(t)=1.96$ for all $t$, this would yield simultaneous coverage lower than $0.95$ due to the multiple comparison problem mentioned earlier. In addition to the aforementioned global control of the coverage probability, a unique feature of the simultaneous confidence bands is that they also allow to draw local inference, e.g. giving at which significance level one can reject the null hypothesis of no effect within a sub-interval of $[0,1]$ \citep[][Prop. 3.2]{Liebl}.

In practice, the covariance structure of the random errors functions (and in extension, $C_\theta$) is likely unknown. However, we can use the residuals from the two outcome regression fits as estimates of the error functions, and their sample covariance function as a consistent estimate $\hat{\sigma}_z(s,t)$ of ${\sigma}_z(s,t)$ to obtain:
\begin{equation}
\label{hatctheta}
    \hat{C}_\theta(s,t)=\hat\sigma_1(s,t) \ \overline{\boldsymbol{x}}\left(\mathbf{X}^T_{1} \mathbf{X}_{1}\right)^{-1} \overline{\boldsymbol{x}}^T + \hat\sigma_0(s,t) \ \overline{\boldsymbol{x}}\left(\mathbf{X}^T_{0} \mathbf{X}_{0}\right)^{-1} \overline{\boldsymbol{x}}^T.
\end{equation}
When this estimate is used in finding $u_\alpha(t)$,a t-process can be used as a finite sample correction, see \cite{Liebl}.

\section{The effects of early adult residence location on cumulative lifetime income}\label{APP}

We aim at studying the consequences of the place of labour market entry on lifetime income trajectories. We focus our analysis on the cohort born 1954 in Sweden, using a population-wide and longitudinal database available at the Ume\aa\ SIMSAM Lab \citep{Lind}. This database allows us to follow income trajectories from the cohort's entrance into the labour market until 2017 (age 63). The potential effect of the time (and place) of labour market entrance on incomes has been widely discussed in the Economics literature; see, e.g., \cite{Altonji, Kwon, Oreo2, Raaum, Schwandt, Aslund}. We here follow an operationalisation of (exposure to) treatment similar to \cite{Raaum} and define the location of the entry in the labour market as the place of residence at the age of 20, i.e. in 1974 for our cohort. Thus, we define a binary treatment, where $z=1$ is defined as living at the age of 20 in the local labour markets of the three biggest Swedish cities (Stockholm, Gothenburg or Malmö). The alternative treatment, $z=0$, is defined as living in local labour markets in which the central municipality had less than 50 000 residents at the age of 20. Local labour markets are units defined by patterns of employment and commuting at the municipality level \citep{LA2,LA}.

The outcome of interest, $y(t)$, is the total earned income (in SEK), defined as all taxable income except from capital. This covers wages and income from business/self-employment, unemployment, parental and sick leave benefits, and pensions. The outcome is measured annually, and data is available between 1968-2017. The income is adjusted for inflation to match the monetary value in 2017 and discounted with a factor of 0.03 \citep{disk}. We calculate the cumulative income over the period 1975-2017, i.e., starting from the year following treatment exposure. Lastly, this cumulative income is logarithmized, a common transformation for incomes that normalises their distribution. In the sequel, "income" will refer to the logarithmized cumulative income.

We control for a number of pre-treatment covariates: \textit{Year of first income}, the year in which an individual first received income if this occurs before exposure to the treatment (before 1974); \textit{Previous income}, which is the logarithmized cumulative income an individual has accumulated before exposure; \textit{Children before age 20}, a binary indicator of the subject having at least one child before exposure; and an indicator of completing (at least) upper \textit{secondary education}. The latter is based on the subjects' highest achieved education level in 1990 as a proxy, as this is the first year for which the variable is available annually. In addition, we control for some family measures: The subjects' \textit{number of siblings} (in 1973); \textit{Parents' income} during the subjects' adolescence (the logarithmized total for both parents for the years 1968-1971); and an indicator indicating if both parents are born outside of Sweden (\textit{Foreign-born parents}). We also control for the highest education level (lower secondary, upper secondary or tertiary) achieved by either mother, father, or both, in 1970, measured by the indicators of \textit{parents' secondary education} and \textit{parents' tertiary education}. Descriptive statistics for the confounders are shown in Table \ref{descriptives} in Appendix \ref{comparebands}.

The 1954 birth cohort was chosen to give the longest possible period in which income data is available. We limit the analysis to individuals born in Sweden and excluded a total of around 4000 individuals due to missingness on one or more of the covariates, primarily those related to parental background. The resulting data consist of 54485 individuals: 27805 men (55.6\% "exposed", $z=1$) and 26680 women (59.3\% "exposed", $z=1$). In our analyses, we treat men and women as separate populations and target two parameters of interest: the FATE for men, $\theta_M$, and for women, $\theta_W$. After some initial pre-processing, we estimate $\theta_M$ and $\theta_W$ with \eqref{theta} and construct 95\% confidence bands as described in Section \ref{estimation.sec}. Appendix \ref{pseudoalgo.sec} contains further details of this process. All analyses are performed in \texttt{R} (Version 4.0.2) \citep{R}, using the \texttt{ffscb} package \citep{ffscb} to obtain the confidence bands. 

\begin{figure}[h!]
\includegraphics[width=0.9\linewidth]{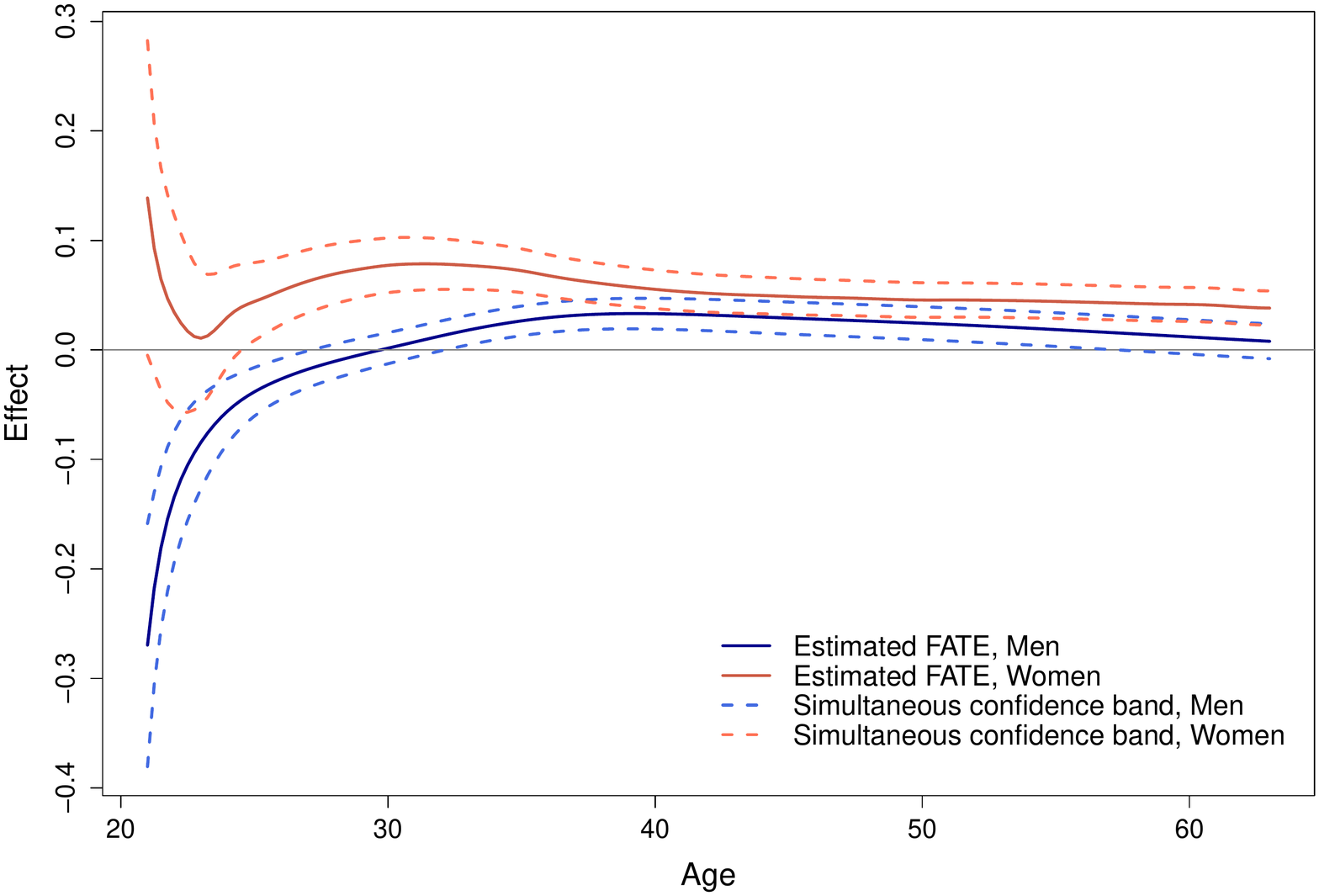} 
\caption{Estimates (solid lines) of the FATE of living in urban versus rural areas at the age of 20 on cumulative life income for the cohort born 1954 in Sweden. 95\% simultaneous confidence bands (dashed lines) for the FATE.  Results from the population of men are shown in blue and results for women in red.}
\label{fig:Confidence bands}
\end{figure}

Figure \ref{fig:Confidence bands} shows the estimates of the two FATE $\theta_W$ and $\theta_M$ and their corresponding 95\% simultaneous confidence bands from the separate analyses in the populations of men (blue) and women (red). Especially in the early years, the two effects display different patterns between the two populations. A caveat to this comparability and to the results discussion below is the possible existence of unobserved confounding. For men, we observe a negative effect of living in an urban area at the age of 20 on early incomes, between the ages of 21 and 26 (1975-1980). This might be due to differences in later educational attainment, i.e. that those who live in an urban area in early adulthood are more likely to continue tertiary education. From around age 30 and onward, the effect for men is positive, and the 95\% simultaneous confidence bands do not contain zero for the majority of these years.

For women, the effect is positive for the entire period, and the bands do not contain zero from 1979 onward. This suggests that women already at an early stage in adulthood tend to benefit from living in an urban location at age 20, and continue to do so for most of their working life. Notice from Figure \ref{fig:Confidence bands} that in the first years after (potential) labour market entry, the confidence bands are wider to account for the larger variability in cumulative incomes during that period. 
Additional figures comparing the simultaneous confidence bands to pointwise bands can be found in Appendix \ref{comparebands}.

\section{Simulation study}\label{SIM}

This section evaluates the finite sample performance of the confidence bands introduced in this paper using Monte Carlo simulations. A realistic setting is obtained by sampling covariate values from the sample used in the case study of Section \ref{APP} and simulating exposure and outcome using the fitted propensity score and outcome models.

\subsection{Study design}
The design described below is replicated for the women and men samples separately. We sample $n$ vector values,  $\boldsymbol{x}_i$, $i=1,\ldots,n$, with replacement from the sample used in Section \ref{APP} to form a simulation replicate of the original sample. We use stratified sampling since some covariate value combinations have few observations. The three strata used are "Children before age 20", "Foreign-born parents", and "Parents’ tertiary education". These simulated values of $\boldsymbol{x}$ are treated as fixed for all replicates in a fixed design setting. A random design setting will also be considered later on.

We generate the treatment assignments as $z_i \sim Ber(\pi_i)$, with 
\begin{equation}
\label{pshat}
    \pi_{i} = \Pr(z_i = 1 \mid \boldsymbol{x}_i) = \frac{exp(\boldsymbol{x}_i^{T} \boldsymbol{\gamma})}{1+exp(\boldsymbol{x}_i^{T} \boldsymbol{\gamma})}, 
\end{equation}
where $\boldsymbol{\gamma}$ is based on estimates from the application study and given in Table \ref{gamma} in Appendix \ref{simdetails.sec}. 
The potential outcomes are generated at 43 equidistant points $t$ on [0,1] as
\begin{equation}\label{flmigen}
y_i(t) = \left\{
\begin{array}{rl}
\boldsymbol{x}^{T}_i  \boldsymbol{\beta}_0(t) + \varepsilon_{0i}(t) & ,\ \text{for } i: z_i = 0,\\
\boldsymbol{x}^{T}_i  \boldsymbol{\beta}_1(t) + \varepsilon_{1i}(t) & ,\ \text{for } i: z_i = 1,
\end{array} \right.
\end{equation}
where $\boldsymbol{\beta}_z(t)$, $z=0,1$, are based on the estimated regression coefficients from the corresponding models in the application study.
They are displayed in Figure \ref{fig:betahat} in the Appendix (exact numerical values can be obtained from the authors). Moreover, 
the random error functions $\varepsilon_{zi}(t)$ are generated such that
\begin{equation}
    \varepsilon_{zi} \sim \mathcal{GP}(0, \sigma_z),
    \label{sigma_sim}
\end{equation}
where $\sigma_z$ are calculated from the residuals of the outcome regression models in Section \ref{APP}.
In the fixed design setting, the \textit{"true"} FATE parameter is $\overline{\boldsymbol{x}}^{T}  \boldsymbol{\beta}_1(t) - \overline{\boldsymbol{x}}^{T}  \boldsymbol{\beta}_0(t)$,
where 
$    \overline{\boldsymbol{x}} = \frac{1}{n}\, \boldsymbol{x}^T \mathbb{1}_{n \times 1}.
    \label{xbar*}
$

We use 1000 replicates to evaluate four different types of pointwise and simultaneous confidence bands for $\hat{\theta}(t)$:

\textit{(i)} We assume that the covariance structures $\sigma_0$ and $\sigma_1$, used to generate the random error functions in \eqref{sigma_sim}, are known. These covariances are used to calculate $C_\theta(s,t)$. We then construct 95\% simultaneous confidence bands as in \eqref{band}.

\textit{(ii)} We estimate the error functions' covariance structures $\sigma_0$ and $\sigma_1$ using the residuals from the regression models in the current simulation replicate. This estimated covariance structure is then used to construct the confidence bands. 

\textit{(iii)} We construct pointwise 95\% confidence intervals separately at each of the 43 time points for which the outcome is observed. For this third type of band, we use the diagonal of the known true covariance matrix $C_\theta(s,t)$ to construct the pointwise 95\% confidence band: 
\begin{equation}
    \label{t interval1}
    \hat{\theta}(t) \ \pm \ \ t_{.05} \sqrt{C_\theta(t,t)},
\end{equation}
where $t_{.05}$ is the 97.5th percentile from the t-distribution with $n-1$ degrees of freedom.

\textit{(iv)} We construct the pointwise confidence bands with \eqref{t interval1}, but using the estimator $\hat{C}_\theta(t,t)$ in \eqref{hatctheta} instead of $C_\theta(t,t)$. 

The empirical coverage rate is computed as $100$ times the proportion of replicates for which the band contains $\theta(t)$ over all observed time points $t$. Four different sample sizes are used, $n=250, 500, 1000$ and $10000$.

Aside from the above setup using a fixed design and Gaussian random errors, we also compare the bands of type \textit{(ii)} and \textit{(iv)} in two other setups:  Firstly, we perform the simulations in a random design setup, in which new covariate values are re-sampled for each replicate. For the evaluation of the confidence bands, we use the approximation of the "true" FATE parameter:
\begin{equation*}
    \frac{1}{1000} \sum_{b=1}^{1000} \left(\overline{\boldsymbol{x}}^{T}_b  \boldsymbol{\beta}_1(t) - \overline{\boldsymbol{x}}^{T}_b  \boldsymbol{\beta}_0(t)\right),
    \label{theta 0 random}
\end{equation*}
where $\boldsymbol{x}_b$ is the covariate values in the $b^{th}$ replicate. 

Secondly, for the fixed design setup, we allow in another scenario for an error process which is not Gaussian: instead of \eqref{sigma_sim}, we generate the error term from a multivariate t-distribution with ten degrees of freedom \citep{mvt}:  
$
\varepsilon_{zi}(t) \sim t_{10}\left(\boldsymbol{0}, \sigma_z(t,t)\right).
$

\subsection{Results}

Table 1 displays empirical coverage rates for the four confidence bands studied obtained over the 1000 replicates in the fixed design setup. There are no qualitative differences in the results between the simulations based on the data for men and women. The simultaneous confidence bands using the known covariance (type \textit{i}) attain their nominal size of 95\%. When estimating the covariance function $\hat{C}_\theta(s,t)$ (bands of type \textit{ii}), we see that empirical coverages attain their nominal level of 95\% for sample sizes $n=500$ and larger. As expected, both types of pointwise confidence bands (type \textit{iii} and \textit{iv}) have too low empirical coverages due to the multiple comparison problem.

\begin{figure}[h!t]
\includegraphics[width=0.9\linewidth]{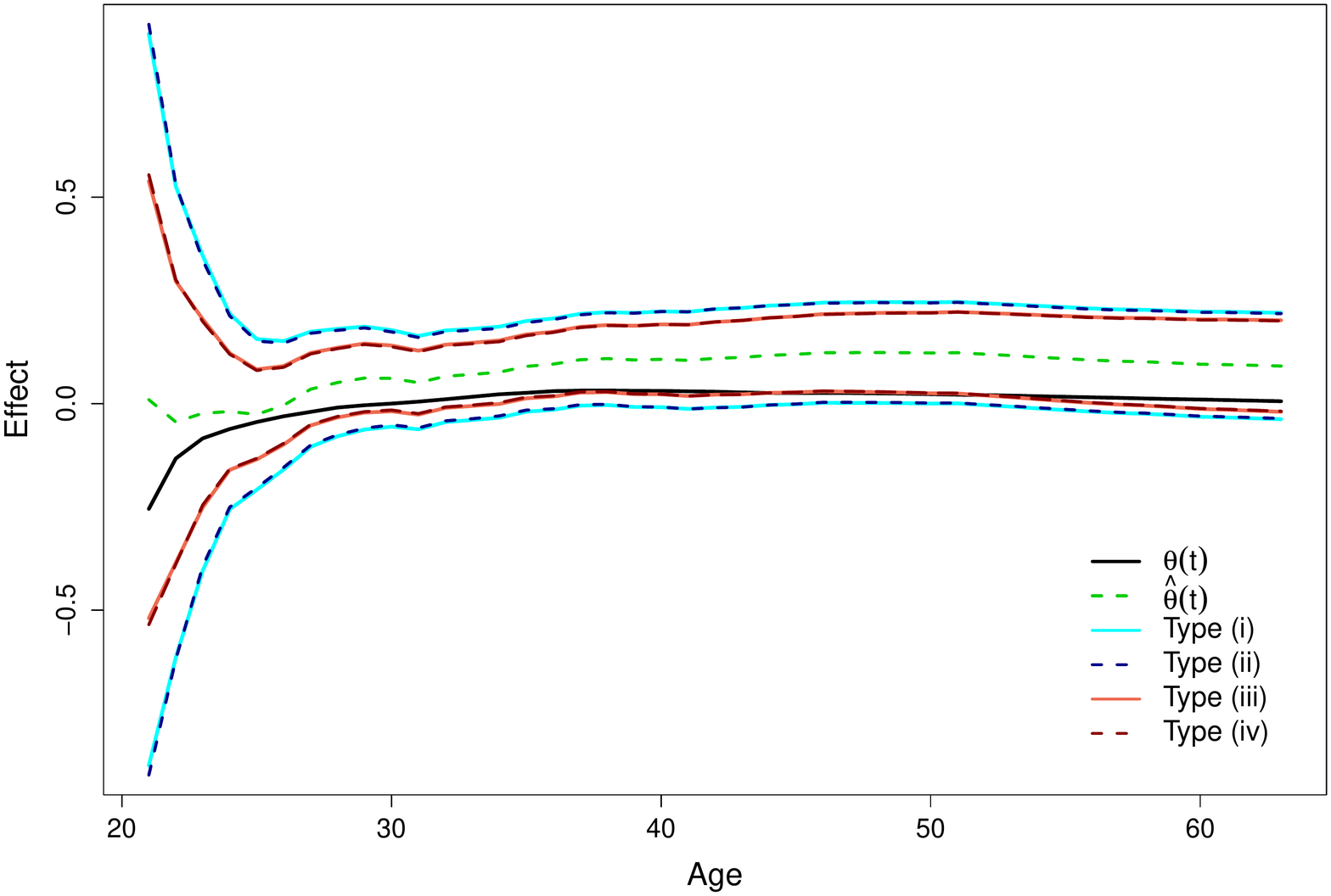} 
\caption{True and estimated FATE and the four types of confidence bands from one simulated replicate for the fixed design and male subset, n=500.}
\label{fig:Simulation}
\end{figure}

Figure \ref{fig:Simulation} displays the results graphically from one simulation replicate with $n=500$ using the fixed design and the male subset. The simultaneous confidence bands (type \textit{i} and \textit{ii}), shown as dotted and dashed red curves are wider than their pointwise counterparts and cover the true $\theta(t)$ at all points.  An additional figure displaying the average estimates and confidence bands over these 1000 replicates can be found in Appendix \ref{Simcurves}.

\begin{center}
\begin{table}\label{simresults}
\caption{Empirical coverages based on 1000 replicates for four types of confidence bands in the fixed design, four sample sizes, presented for men and women separately. The nominal coverage level is 95\%.}
\begin{tabular}{l c c c c  c c c c c}
  \hline
  \hline
 Fixed design & \multicolumn{4}{c}{Men} & &\multicolumn{4}{c}{Women}\\
  Type\textbackslash $n$ & 250 & 500 & 1000 & 10000 && 250 & 500 & 1000 & 10000 \\
  \hline
  \textit{(i)} & 95.0 & 96.4 & 95.7 & 95.2  && 95.2 & 96.0 & 95.9 & 95.7\\
  \textit{(ii)} & 93.4 & 96.1 & 95.5 & 95.1 && 93.2 & 95.4 & 95.8 & 95.6\\
  \textit{(iii)} & 70.6 & 69.7 & 70.9 & 72.1 && 71.0 & 71.2 & 72.7 & 70.1 \\
  \textit{(iv)} & 66.2 & 67.5 & 69.7 & 72.1 && 65.5 & 65.5 & 70.8 & 69.9 \\
  \hline
\end{tabular}
\end{table}
\end{center}

The results for the two additional scenarios can be found in Table 2. We can see that the results from the random design setting are similar to those from the fixed design. The bands attain a coverage close to the nominal level of 95\% for the sample sizes considered. In addition, the simultaneous confidence bands also perform well under non-Gaussian random error terms (see rows corresponding to Non-GP errors in Table 2).

\begin{center}
\begin{table}\label{simresults2}
\caption{Empirical coverages based on 1000 replicates for additional scenarios, presented for men and women separately. The nominal coverage level is 95\%.}
\begin{tabular}{l l c c c c  c c c c c}
  \hline
  \hline
   & & \multicolumn{4}{c}{Men} & &\multicolumn{4}{c}{Women}\\
 Setup & Type\textbackslash  $n$ & 250 & 500 & 1000 & 10000 && 250 & 500 & 1000 & 10000 \\
  \hline
 Random design & \textit{(ii)} & 93.8 & 95.3 & 94.6 & 96.9 && 95.5 & 94.6 & 95.6 & 96.3\\
  & \textit{(iv)} & 67.3 & 70.3 & 72.4 & 72.9 && 67.0 & 67.3 & 69.0 & 69.9 \\
  \hline
 Non-GP errors & \textit{(ii)} & 93.4 & 95.3 & 96.0 & 95.2 && 95.2 & 96.2 & 95.4 & 96.1\\
  & \textit{(iv)} & 66.9 & 69.2 & 70.9 & 71.5 && 63.9 & 67.3 & 68.0 & 72.3 \\
  \hline
\end{tabular}
\end{table}
\end{center}

\section{Discussion}
This paper presents methods to study causal effects of a treatment on a functional outcome. In this context, we define a functional causal parameter (FATE or FATT), propose an estimator based on outcome regression, and provide simultaneous confidence bands that cover the true functional parameter over its entire domain with a given probability. The theoretical results are confirmed in simulation experiments, where the simultaneous confidence bands achieve the desired nominal coverage rate.

The methods derived here have allowed us to study the effect of early adult residence location on cumulative lifetime incomes over the entire working life of a Swedish birth cohort. For men, we find a negative effect of living in an urban versus a rural area at the age of 20 on cumulative incomes in early career stages. This effect becomes positive later in life. For women, living in an urban area in early adulthood has a slightly stronger, positive effect on cumulative incomes throughout the working life. The results of such an observational study must be taken with caution, in particular because of the possibility of existing unobserved confounding (Assumption 2 violated). 
On the other hand, earlier observational studies on the associations between initial labour market conditions and later labour market outcomes have used much shorter follow-up periods \citep{Altonji, Kwon, Oreo2, Schwandt,Aslund}, or do not correct for the multiple comparison problem \citep{Raaum}.
 Note that the methods and theory presented are readily applicable to other contexts. For instance, patients may be monitored with electronic instruments yielding functional signal outcomes in medical applications. 
 
 We have focused on outcome regression estimators and only allow for scalar covariates. Theoretical results for other estimators of average causal effects, including the augmented inverse probability weighting estimator and the closely related targeted learning estimator \citep{tmle}, as well as to allow for functional covariates would be valuable future research directions.
  In all the above cases, the difficulty resides in showing that the resulting functional estimator is asymptotically a Gaussian or elliptical process. In particular, the tightness of the pointwise estimators needs to be shown, which is known to be problematic with functional covariates \citep{Choi}.

\section*{Acknowledgements}
We are grateful to Xijia Liu for comments that have improved the paper. 

\section*{Declarations of interest}
The authors declare no competing interests.

\section*{Funding sources}
The work was supported by the Swedish Research council [grant number 2016-02851]. The
Umeå SIMSAM Lab data infrastructure in Umeå used in this study was developed with
support from the Swedish Research Council [grant number 2008-7491], the Riksbanken
Jubileumsfond and by strategic funds from Umeå University.

\section*{Data availability statement}
The data analysis is based on a record linked register database available at the Umeå SIMSAM Laboratory (https://www.umu.se/en/research/infrastructure/umea-simsam-lab/) at Umeå University, Sweden. The database is built as a combination of different population-based registers linked through Swedish personal registration numbers and was compiled in collaboration with different Swedish authorities. Both the approval from the Ethical vetting board and the contracts we have signed with the Swedish authorities do not allow us to give away the data to a third party. The data can, however, be accessed by researchers wanting to replicate the analysis, although this can be done only locally at the Umeå SIMSAM Laboratory, where the data is stored on servers disconnected from the internet.

\bibliography{sn-bibliography}

\section*{Figure legends}
Figure 1: Mean differences in LCI (urban versus rural location at age 20), together with 95\% pointwise confidence bands, for the 1954 birth cohort in Sweden. By gender, blue: men, red: women.

Figure 2: Estimates (solid lines) of the FATE of living in urban versus rural areas at the age of 20 on cumulative life income for the cohort born 1954 in Sweden. 95\% simultaneous confidence bands (dashed lines) for the FATE.  Results from the population of men are shown in blue and results for women in red.

Figure 3: True and estimated FATE and the four types of confidence bands from one simulated replicate for the fixed design and male subset, n=500.

Figure D1: Histogram of estimated propensity scores for the treated and controls among men.

Figure D2: Histogram of estimated propensity scores for the treated and controls among women.

Figure D3: Estimates (solid line) of the FATE of living in urban versus rural areas at the age of 20 on cumulative life income for the male cohort born 1954 in Sweden. 95\% simultaneous confidence bands (dashed line) and 95\% pointwise confidence bands (dotted line) for the FATE.

Figure D4: Estimates (solid line) of the FATE of living in urban versus rural areas at the age of 20 on cumulative life income for the female cohort born 1954 in Sweden. 95\% simultaneous confidence bands (dashed line) and 95\% pointwise confidence bands (dotted line) for the FATE.

Figure E1: Estimated regression coefficients for the two treatment groups, for men and women.

Figure E2: Means of the FATE estimates and four types of confidence bands over the 1000 simulation replicates for male subset, n = 500.

\newpage

\begin{appendix}
\counterwithin{figure}{section}
\counterwithin{table}{section}
APPENDIX
\section{Functional Average Treatment effect on the Treated population}\label{fatt.sec}

The Functional Average Treatment effect on the Treated population (FATT) is defined as:
\begin{equation*}
  \eta := \left\{\eta(t) =\mathbb{E}\left( y_{1i}(t)-y_{0i}(t) \mid  z_i=1 \right):t \in [0,1]\right\}.  
\end{equation*}

\noindent It is identified under Assumptions 1.a and 2.a in section \ref{identification.sec}.
Under model \eqref{flm}, an estimator for the FATT is given by:

\begin{equation*}
    \label{eta_hat2}
    \begin{split}
      \hat{\eta}(t) & =  \frac{1}{n_1} \sum_{i=1}^n z_i \ \Big(\boldsymbol{x}^T_i \ \hat{\boldsymbol{\beta}}_1(t) - \boldsymbol{x}^T_i \ \hat{\boldsymbol{\beta}}_0(t) \Big) \\
      & = \overline{\boldsymbol{x}}_1 \ \hat{\boldsymbol{\beta}}_1(t) - \overline{\boldsymbol{x}}_1 \ \hat{\boldsymbol{\beta}}_0(t),
    \end{split}
\end{equation*}
where $
    \overline{\boldsymbol{x}}_1 = \frac{1}{n_1} \ \mathbb{1}_{1 \times n_1} \mathbf{X}_1
    $
is the vector of sample means for the subset of treated observations. 
Analogous to equations \eqref{dist2}-\eqref{covtheta1}, we have that $\hat{\eta} = \{\hat{\eta}(t): t\in [0,1]\}$ is a Gaussian process:

\begin{equation*}
    \hat{\eta} \sim \mathcal{GP} \left( \overline{\boldsymbol{x}}_1 \ \boldsymbol{\beta}_1 - \overline{\boldsymbol{x}}_1 \ \boldsymbol{\beta}_0, C_\eta \right), 
\end{equation*}
where
\begin{equation*}
    C_\eta (s,t) = \sigma_1(s,t) \ \overline{\boldsymbol{x}}_1\left(\mathbf{X}^T_{1} \mathbf{X}_{1}\right)^{-1} \overline{\boldsymbol{x}}_1^T + \sigma_0(s,t) \ \overline{\boldsymbol{x}}_1\left(\mathbf{X}^T_{0} \mathbf{X}_{0}\right)^{-1} \overline{\boldsymbol{x}}_1^T.
\end{equation*}

\newpage
\section{Estimator of FATE is a Gaussian process}\label{GPproof.sec}

We reproduce here the definition of a multivariate Gaussian process given in \cite{Chen2}:

Let $\boldsymbol{f}$ be a multivariate Gaussian process on $\mathcal{X}$ with vector-valued mean function $\boldsymbol{\mu} : \mathcal{X} \mapsto \mathbb{R}^n$, covariance function $\sigma : \mathcal{X} \times \mathcal{X} \mapsto \mathbb{R}$, and positive semi-definite parameter matrix $\Omega \in \mathbb{R}^{n \times n}$. We denote this process $\boldsymbol{f} \sim \mathcal{MGP}(\boldsymbol{\mu}, \sigma, \Omega)$. Then the vectorization of any finite collection of $L$ vector-valued variables have a joint multi-variate Gaussian distribution:

\begin{equation*}
    vec\left([\boldsymbol{f}(x_1)^T, \dots, \boldsymbol{f}(x_L)^T]^T\right) \sim \mathcal{N}_{dn} \left(vec(M), \Sigma \otimes \Omega\right), L \in \mathbb{N}, 
\end{equation*}
where $M \in \mathbb{R}^{n \times L}$ with $M_{ij} = \mu_j(x_i)$ and $\Sigma > 0, \ \Sigma \in \mathbb{R}^{d \times d}$ with $\Sigma_{ij} = \sigma(x_i, x_j)$. Furthermore, $\boldsymbol{f}, \boldsymbol{\mu} \in \mathbb{R}^n$ are row vectors whose components are the functions $\{ f_i \}_{i = 1}^n$ and $\{\mu_i \}_{i = 1}^n$, respectively. Sometimes $\Sigma$ is called the column covariance matrix, while $\Omega$ is the row covariance matrix.  See also \cite{Chen} for proof of existence and further properties. This in turn implies that any finite collection of $L$ vector-valued variables have a joint matrix variate Gaussian distribution \citep{Gupta}:

\begin{equation*}
    [\boldsymbol{f}(x_1)^T, \dots, \boldsymbol{f}(x_L)^T]^T \sim \mathcal{N}_{d,n} (M, \Sigma \otimes \Omega). 
\end{equation*}

\textit{Proof of Theorem 1:}

Assumption 3 states that the error terms in model \eqref{flm} are multivariate Gaussian processes, denoted as 
\begin{equation*}
     \boldsymbol{\varepsilon}_z \sim \mathcal{MGP}\left( 0, \sigma_z, \mathbb{I}_{n_z} \right),
\end{equation*}
where $\mathbb{I}_{n_z}$ is a $n_z \times n_z$ identity matrix. 
Using the definition above we have that, for any finite collection $\{t_l\}_{l=1}^L$ of points in the domain, the error terms for each treatment group follow a joint matrix-variate Gaussian distribution:  

\begin{equation*}
    \mathrm{E}_z := \left[\boldsymbol{\varepsilon}_z(t_1)^T, \dots, \boldsymbol{\varepsilon}_z(t_L)^T \right] \sim \mathcal{N}_{n_z,L}\left(0_{n_z, L}, \mathbb{I}_{n_z} \otimes \Sigma_z \right),
\end{equation*}
where $\boldsymbol{\varepsilon}_z(t_l), \,l=1, \dots, L,$ are row vectors of length $n_z$; $0_{n_z, L}$ is a $n_z \times L$ null matrix, and $\Sigma_z$ is a $L \times L$ matrix with elements $[\Sigma_z]_{i,j} = \sigma_z(t_i,t_j)$.

We further specify the following matrices: $\Omega_z := (\mathbf{X}^T_{z}\mathbf{X}_z)^{-1}$, $\mathrm{B}_z := \left[\boldsymbol{\beta}_z(t_1), \dots, \boldsymbol{\beta}_z(t_L) \right]$, and $\hat{\mathrm{B}}_z := \left[\hat{\boldsymbol{\beta}}_z(t_1), \dots, \hat{\boldsymbol{\beta}}_z(t_L) \right]$. Given a fixed design matrix, the matrix of estimated regression coefficients can be expressed as

\begin{equation*}
    \hat{\mathrm{B}}_z = \mathrm{B}_z + \Omega_z \mathbf{X}^T_{z} \mathrm{E}_z.
\end{equation*}

Then, for any finite collection of points, $\hat{\mathrm{B}}_z$ follows a matrix-variate Gaussian distribution \citep{Gupta} with

\begin{equation*}
    \hat{\mathrm{B}}_z \sim \mathcal{N}_{(K+1),L} \left(\Omega_z \mathbf{X}^T_{z} 0_{n_z, L}+\mathrm{B}_z, \left[\Omega_z \mathbf{X}^T_{z} \mathbb{I}_{n_z}\left(\Omega_z \mathbf{X}^T_{z}\right)^T \right] \otimes \Sigma_z \right), 
\end{equation*}
which simplifies to 

\begin{equation}
    \label{hat_B_matrix}
    \hat{\mathrm{B}}_z \sim \mathcal{N}_{(K+1),L} \left(\mathrm{B}_z, \Omega_z \otimes \Sigma_z \right).
\end{equation}
Because the above statement holds for any finite collection $\{t_l\}_{l=1}^L$, the process $\hat{\boldsymbol{\beta}}_z = \{\hat{\boldsymbol{\beta}}_z(t): t \in [0,1] \}$ is a multivariate Gaussian process:

\begin{equation*}
    \hat{\boldsymbol{\beta}}_z \sim \mathcal{MGP}\left( \boldsymbol{\beta}_z, \sigma_z, (\mathbf{X}^T_{z}\mathbf{X}_z)^{-1} \right).
\end{equation*}

To derive the distribution of $\hat{\theta}$, we can use the distribution of $\hat{\mathrm{B}}_z$ in \eqref{hat_B_matrix} and see that the product  

\begin{equation*}
    \overline{\boldsymbol{x}}\, \hat{\mathrm{B}}_z \sim \mathcal{N}_{1,L} \left(\overline{\boldsymbol{x}}\, \mathrm{B}_z, \left[ \overline{\boldsymbol{x}}\, \Omega_z \overline{\boldsymbol{x}}^T \right] \otimes \Sigma_z \right).
\end{equation*}

This $1 \times L$-dimensional matrix-variate Gaussian distribution coincides with a $L$-dimensional multivariate Gaussian distribution. Since these distributional statements, again, hold for any finite collection of points in the domain, it follows that $\overline{\boldsymbol{x}}\, \hat{\boldsymbol{\beta}}_z = \{\overline{\boldsymbol{x}}\, \hat{\boldsymbol{\beta}}_z(t) : t \in [0,1]\}$ is a Gaussian process:

\begin{equation*}
    \overline{\boldsymbol{x}}\, \hat{\boldsymbol{\beta}}_z \sim \mathcal{GP} (\overline{\boldsymbol{x}}\,  \boldsymbol{\beta}_z,  \overline{\boldsymbol{x}} \left(\mathbf{X}^T_z \mathbf{X}_z\right)^{-1} \overline{\boldsymbol{x}}^T\sigma_z).
\end{equation*}

Under the assumption of mean ignorability, $\overline{\boldsymbol{x}}\, \hat{\boldsymbol{\beta}}_0$ and $\overline{\boldsymbol{x}}\, \hat{\boldsymbol{\beta}}_1$ are independent, meaning that the difference between them also is a Gaussian process:

\begin{equation*}
    \hat{\theta} \sim \mathcal{GP} \left( \overline{\boldsymbol{x}}\, \boldsymbol{\beta}_1 - \overline{\boldsymbol{x}}\, \boldsymbol{\beta}_0, C_\theta \right), 
\end{equation*}
with $C_\theta$ given in \eqref{covtheta1}.

\

\newpage
\section{Practical details of estimating the FATE and constructing confidence bands}\label{pseudoalgo.sec}

The following section describes the technical details of the application study in section \ref{APP}. All steps were performed in \texttt{R} \citep{R}, and done separately for men and women. 

\textbf{1.} To represent the discrete annual measurements of income as functional objects, we apply a low degree of monotone smoothing to the logarithmized cumulative incomes. The smoothing is performed using the \texttt{smooth.monotone} function in the \texttt{fda} package \citep{fdapack}, and we use a basis of 45 cubic B-splines and a smoothing parameter of $\lambda = 0.1$, penalising the second derivative. During this stage, an additional 28 individuals had to be excluded from the analysis, because the shape of their income accumulations (very steep inclines followed by long plateaus) did not allow for a low degree of monotone smoothing. The resulting functional data objects are evaluated at a quarterly basis, i.e. yielding four evaluations per calendar year. Note that performing the same analyses of the data without or little smoothing yields very similar results.

\textbf{2.} Next, we subset the observed data into two groups: treated ($z=1$) and controls ($z=0$).

\textbf{3.} In each treatment group, we regress the quarterly evaluations of the smoothed functions, $y(t)$, on the covariates $\boldsymbol{x}$, using the \texttt{IWTlm} function in the \texttt{fdatest} package \citep{package}. The estimated regression coefficients $\hat{\boldsymbol{\beta}}_z(t)$ and the residuals from the two outcome regression models are saved. 

\textbf{4.} We then use the estimated regression coefficients and the covariate means to estimate the ATE with $\hat{\theta}(t)$ given in \eqref{theta}.

\textbf{5.} We compute the sample covariances of the residuals from the two outcome regression models in step 2; yielding $\hat{\sigma}_1(s,t)$ and $\hat{\sigma}_0(s,t)$. These are used to compute $\hat{C_{\theta}}$ in \eqref{hatctheta}.

\textbf{6.} With $\hat{C_{\theta}}$, we use the \texttt{cov2tau\_fun} function in the \texttt{ffscb} package \citep{ffscb} to obtain the estimated roughness parameter function $\hat{\tau}$; see \citet{Liebl}.

\textbf{7.} In the same package, we use the \texttt{confidence\_bands} function to construct the simultaneous confidence bands using $\hat{\theta}$, $\hat{C_{\theta}}$ and $\hat{\tau}$. We choose to obtain 95\% simultaneous confidence bands for $\theta(t)$ as well as the corresponding naive point-wise confidence bands based on the t-distribution. All other arguments of this function are set to their defaults.

\newpage
\section{Details of the case study in Section 3}\label{comparebands}

\begin{table}
\caption{Descriptive information on the covariates, presented as mean (standard deviation) or as \% for indicator variables.}
\begin{center}
\begin{tabular}{ l  c  c }
 \hline
 \hline
 Variable & Men & Women \\ 
 \hline
Year of first income & 1971.7 (1.2) & 1972.1 (1.0) \\
Previous income & 8.3 (4.1) & 7.2 (4.5) \\
 Children before age 20 & 1.9\% &  10.0\% \\
 Secondary education & 71.0\% &  79.0\% \\
 Number of siblings & 2.3 (1.8) & 2.3 (1.8) \\
 Parents' income & 11.8 (0.9) & 11.9 (0.9) \\
 Foreign-born parents & 2.3\% & 2.3\% \\
 Parents' secondary education & 29.7\% & 30.0\% \\ 
 Parents' tertiary education & 11.2\% & 11.3\% \\
 \hline
 \hline
\end{tabular}
\end{center}
\label{descriptives}
\end{table}

\subsection{Descriptives on the covariates}\label{simdescriptives}
Table \ref{descriptives} shows the distribution of the covariates for men and women separately.

\subsection{Empirical check of the overlap assumption}
\begin{figure}[h!]
    \centering
    \includegraphics[width=0.9\linewidth]{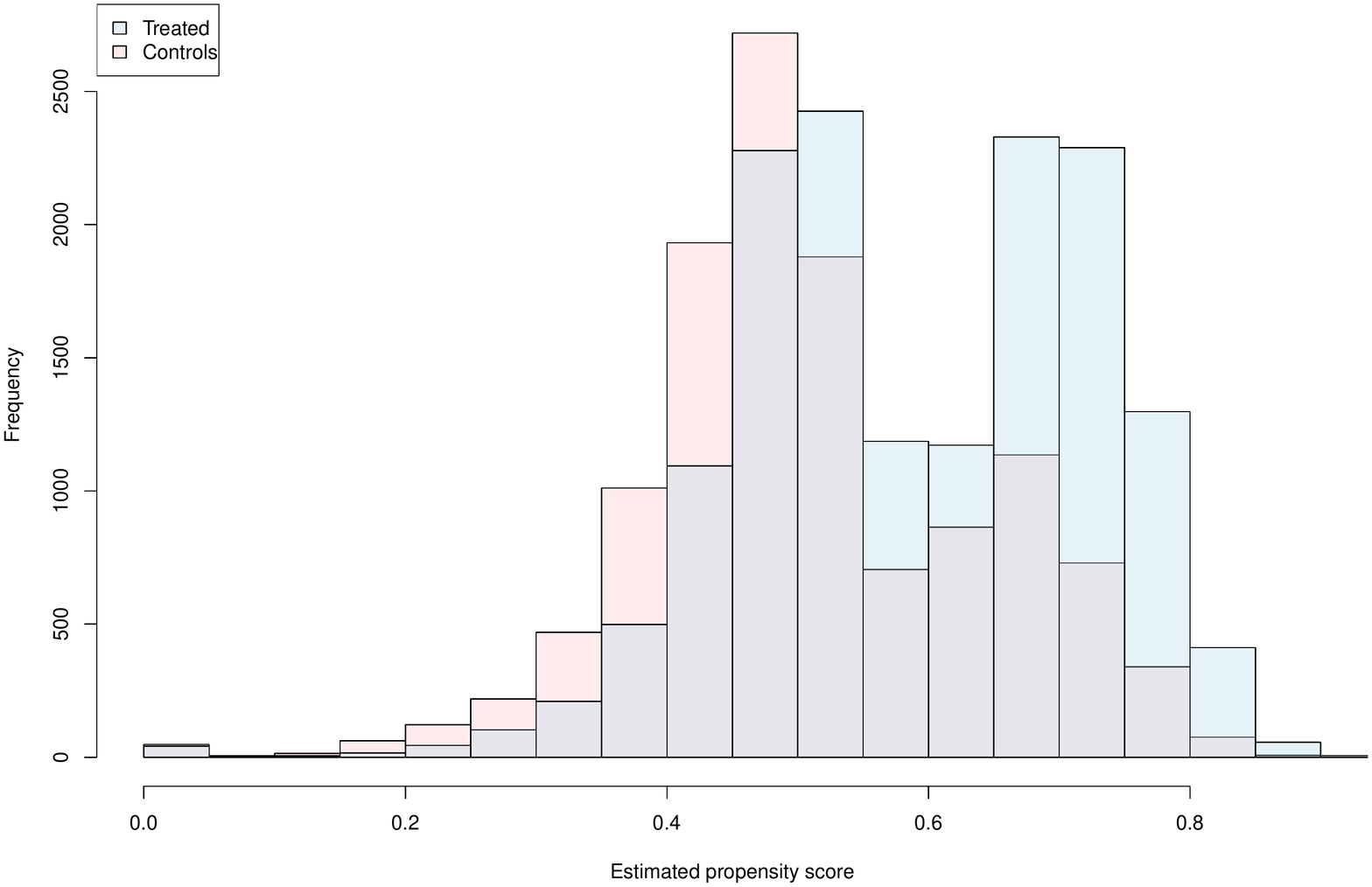}
    \caption{Histogram of estimated propensity scores for the treated and controls among men.}
    \label{fig:overlapM}
\end{figure}

\begin{figure}[h!]
    \centering
    \includegraphics[width=0.9\linewidth]{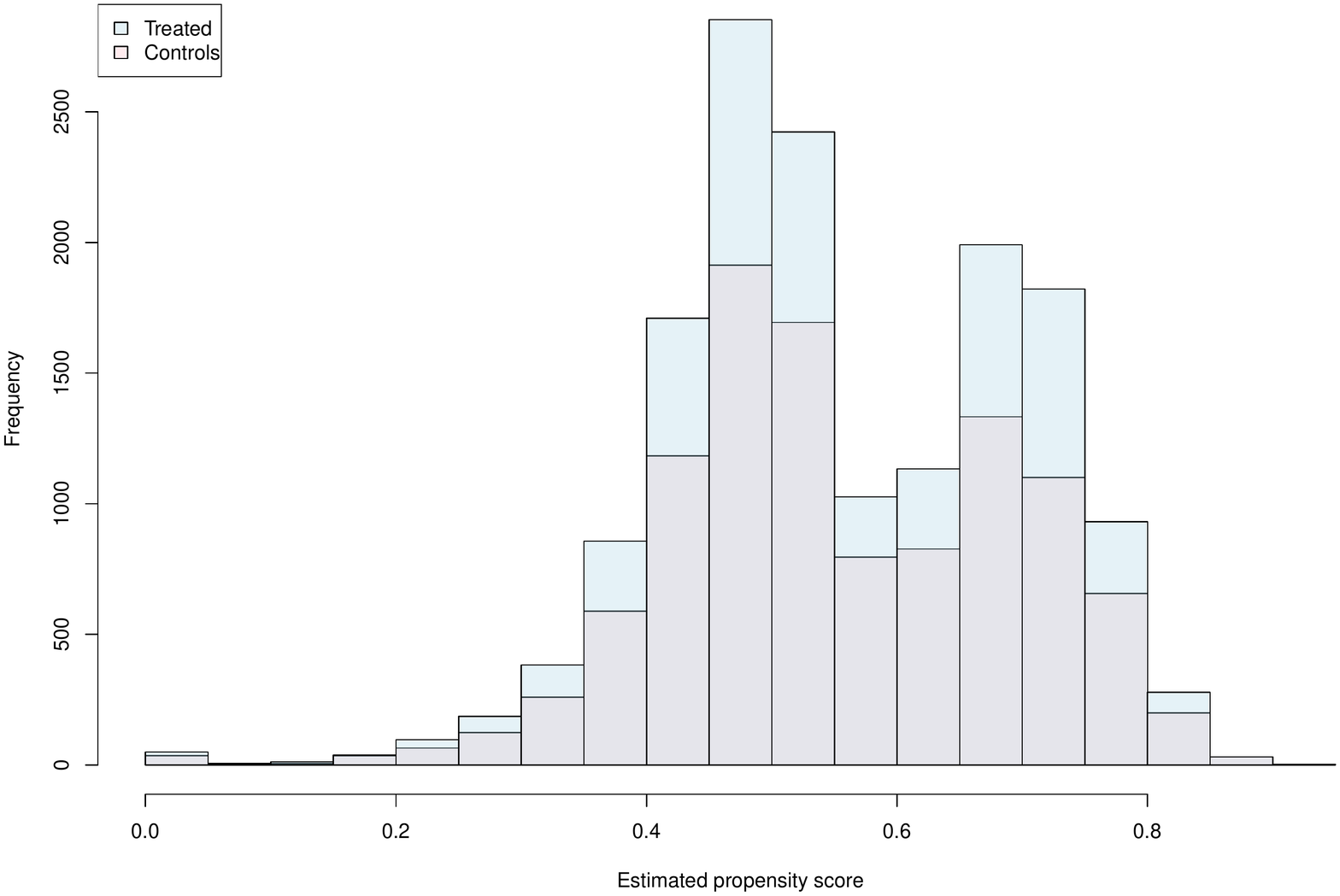}
    \caption{Histogram of estimated propensity scores for the treated and controls among women.}
    \label{fig:overlapW}
\end{figure}

Figures \ref{fig:overlapM} and \ref{fig:overlapW} can be used to investigate Assumption 2 (Overlap) for men and women, respectively. For men, the support of the estimated propensity score ranges from 0.02 to 0.94 among the treated and from 0.01 to 0.91 among the controls. For women, the support of the estimated propensity score ranges from 0.02 to 0.94 among the treated and from 0.03 to 0.92 among the controls. This means that there is no evidence against the overlap assumption.

\subsection{Average estimates and confidence bands}
\begin{figure}[h!]
\includegraphics[width=0.9\linewidth]{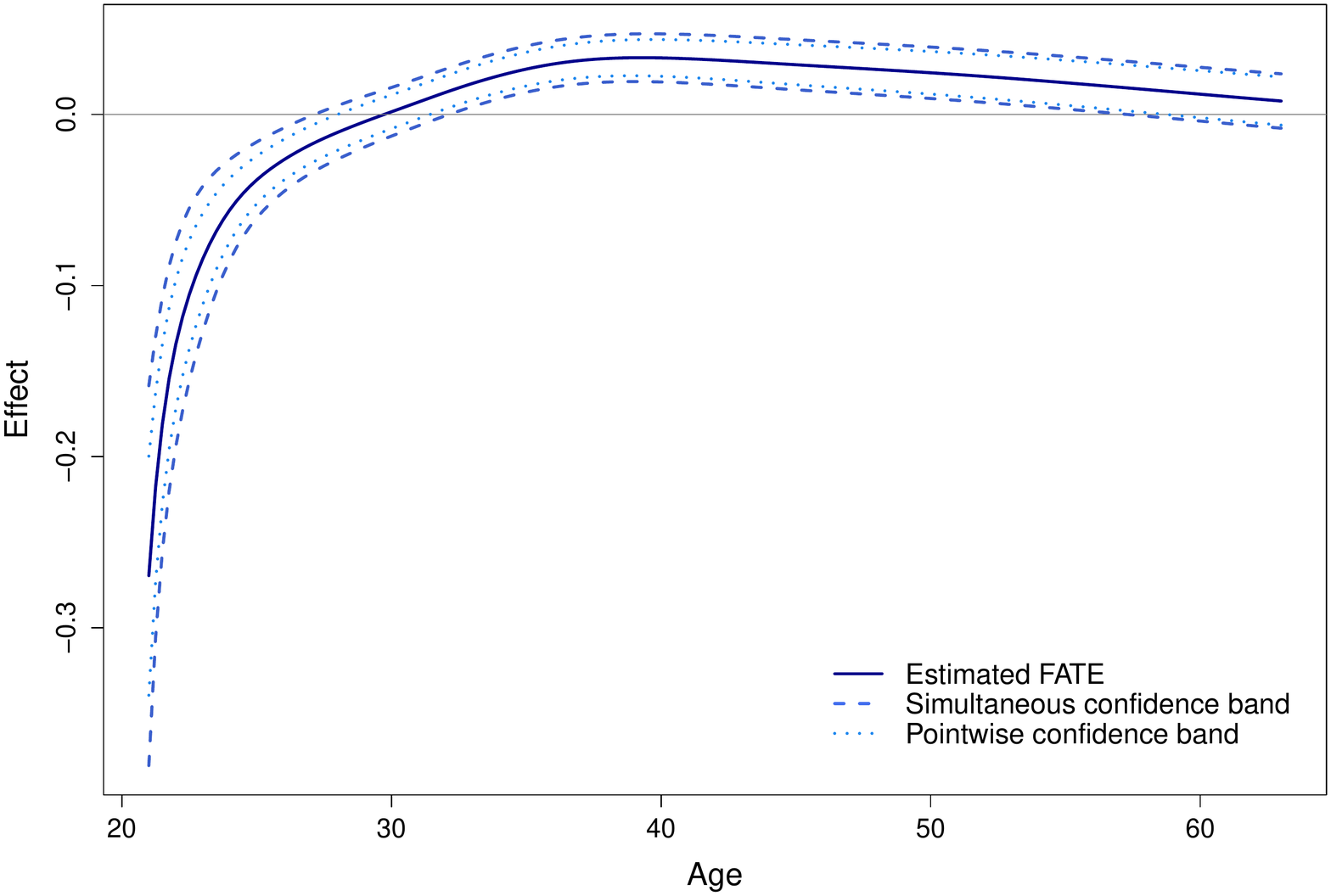} 
\caption{Estimates (solid line) of the FATE of living in urban versus rural areas at the age of 20 on cumulative life income for the male cohort born 1954 in Sweden. 95\% simultaneous confidence bands (dashed line) and 95\% pointwise confidence bands (dotted line) for the FATE.}
\label{fig:compbandsM}
\end{figure}

\begin{figure}[h!]
\includegraphics[width=0.9\linewidth]{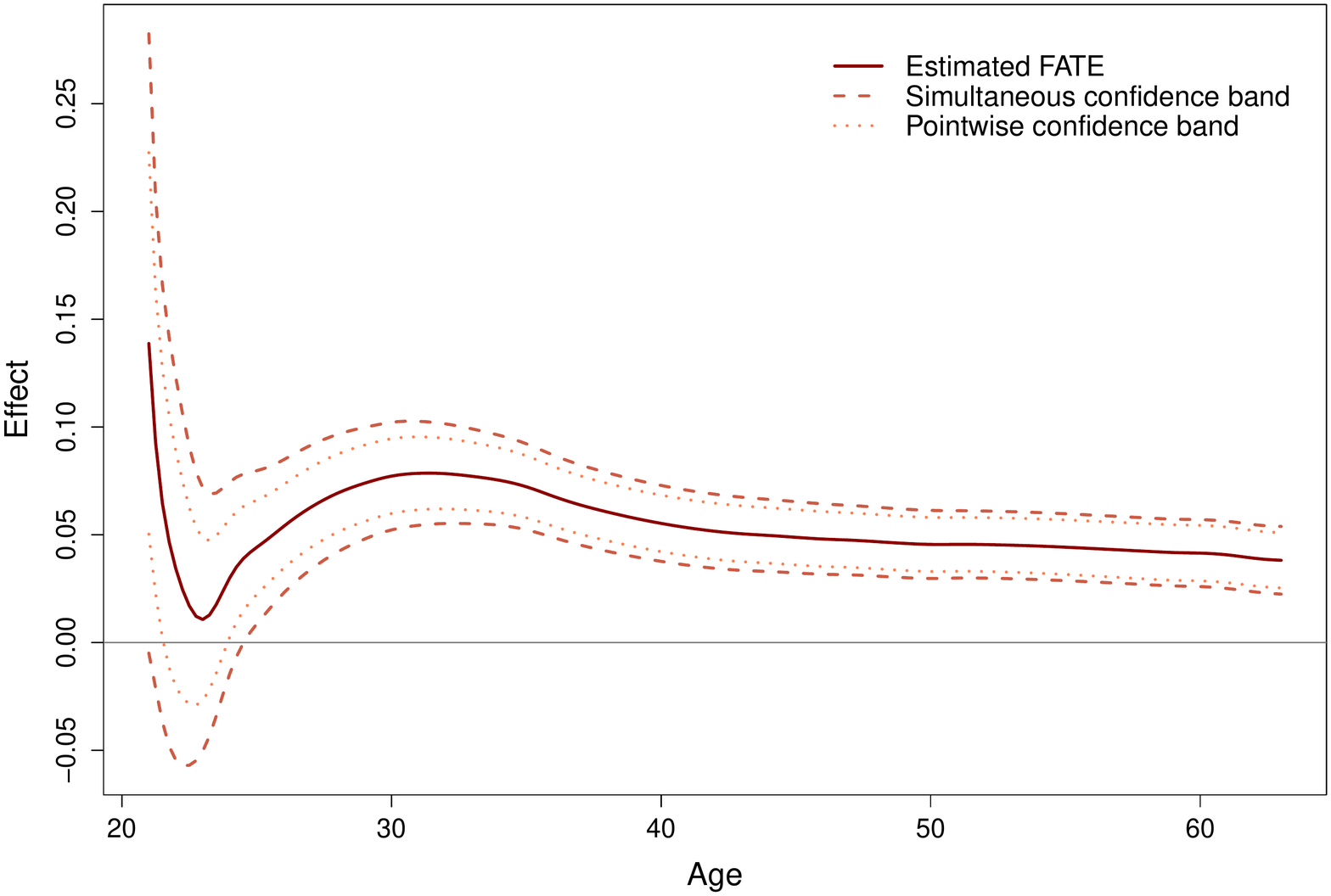} 
\caption{Estimates (solid line) of the FATE of living in urban versus rural areas at the age of 20 on cumulative life income for the female cohort born 1954 in Sweden. 95\% simultaneous confidence bands (dashed line) and 95\% pointwise confidence bands (dotted line) for the FATE.}
\label{fig:compbandsW}
\end{figure}

Figures \ref{fig:compbandsM} and \ref{fig:compbandsW} display 95\% pointwise confidence bands for the FATE estimated in Section 3, in addition to the simultaneous bands, for men and women respectively. We can see that the pointwise bands are more narrow, most notably so for women in the first 10 years. 

\newpage
\section{Details of the simulation study in Section 4}\label{simdetails.sec}

\subsection{Parameters used to generate data}\label{simcoefficients}
\begin{figure}[h!]
\includegraphics[width=0.9\linewidth]{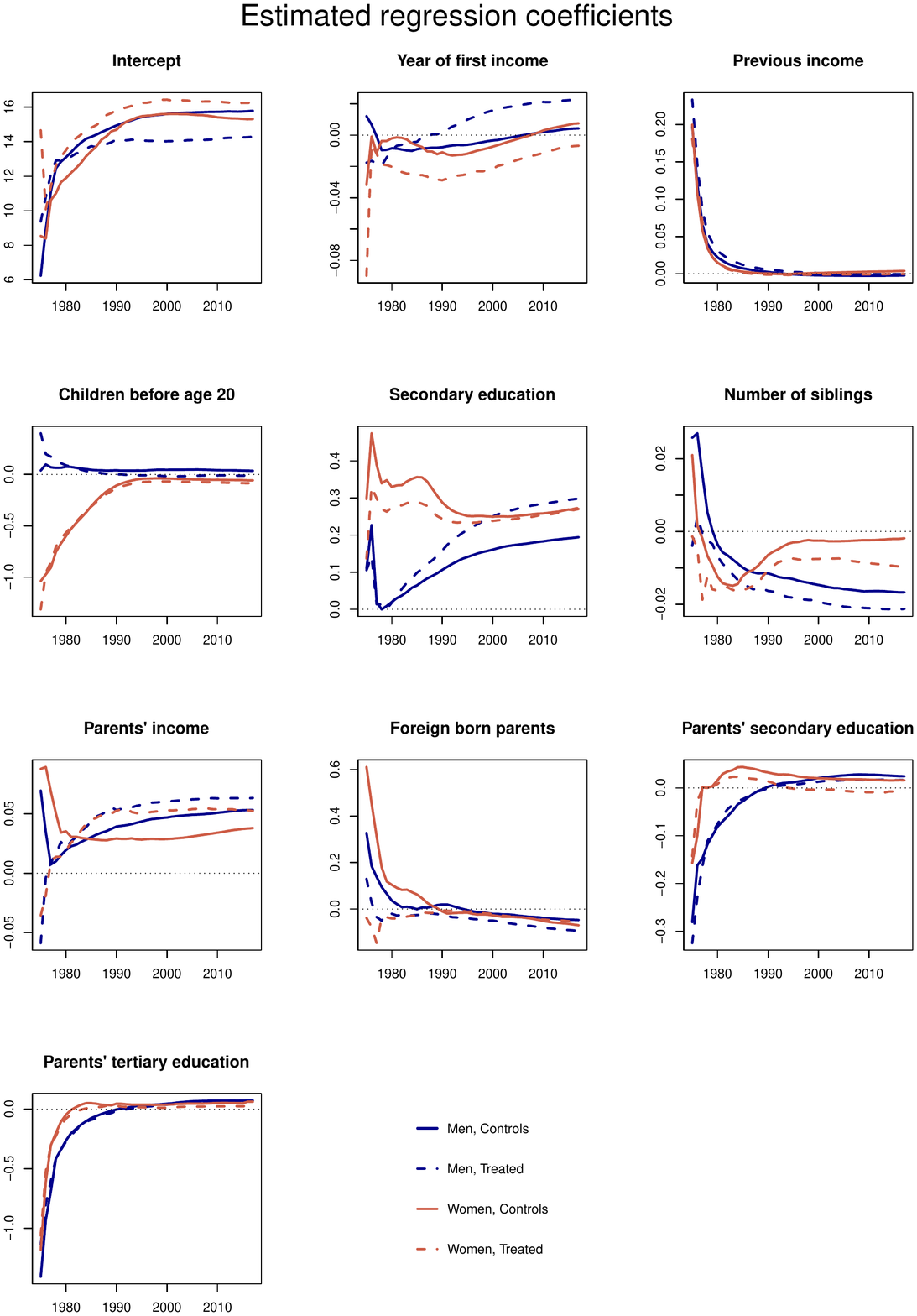} 
\caption{Estimated regression coefficients for the two treatment groups, for men and women.}
\label{fig:betahat}
\end{figure}

The coefficients used in the outcome models \eqref{flmigen} are displayed in Figure \ref{fig:betahat}; exact values can be obtained from the authors. The coefficients used in the propensity score models in \eqref{pshat} are shown in Table \ref{gamma}. 

\begin{table}
\caption{Estimated coefficients ($\boldsymbol{\gamma}$) from the propensity score models (logistic regression). Coefficients in bold are significant at $\alpha=0.05$.}
\begin{center}
\begin{tabular}{ l c  c }
  \hline
  \hline
 Variable & Men & Women \\ 
 \hline
 Intercept & -\textbf{2.194} & \textbf{3.872} \\
 Year of first income & \textbf{-0.052} & \textbf{-0.124} \\
 Previous income & -0.001 & \textbf{0.054} \\
 Children before age 20 & 0.020 & \textbf{-0.378} \\
 Secondary education& \textbf{-0.097} &  \textbf{0.109} \\
 Number of siblings & \textbf{-0.056} & \textbf{-0.052} \\
 Parents' income &\textbf{ 0.510} & \textbf{0.407} \\
 Foreign-born parents & \textbf{0.568} &\textbf{ 0.541} \\
 Parents' secondary education & \textbf{0.674} & \textbf{0.690} \\ 
 Parents' tertiary education & \textbf{0.851} & \textbf{0.933} \\
  \hline
  \hline
\end{tabular}
\end{center}
\label{gamma}
\end{table}

\subsection{Average estimates and confidence bands}\label{Simcurves}

\begin{figure}[h!]
\includegraphics[width=0.9\linewidth]{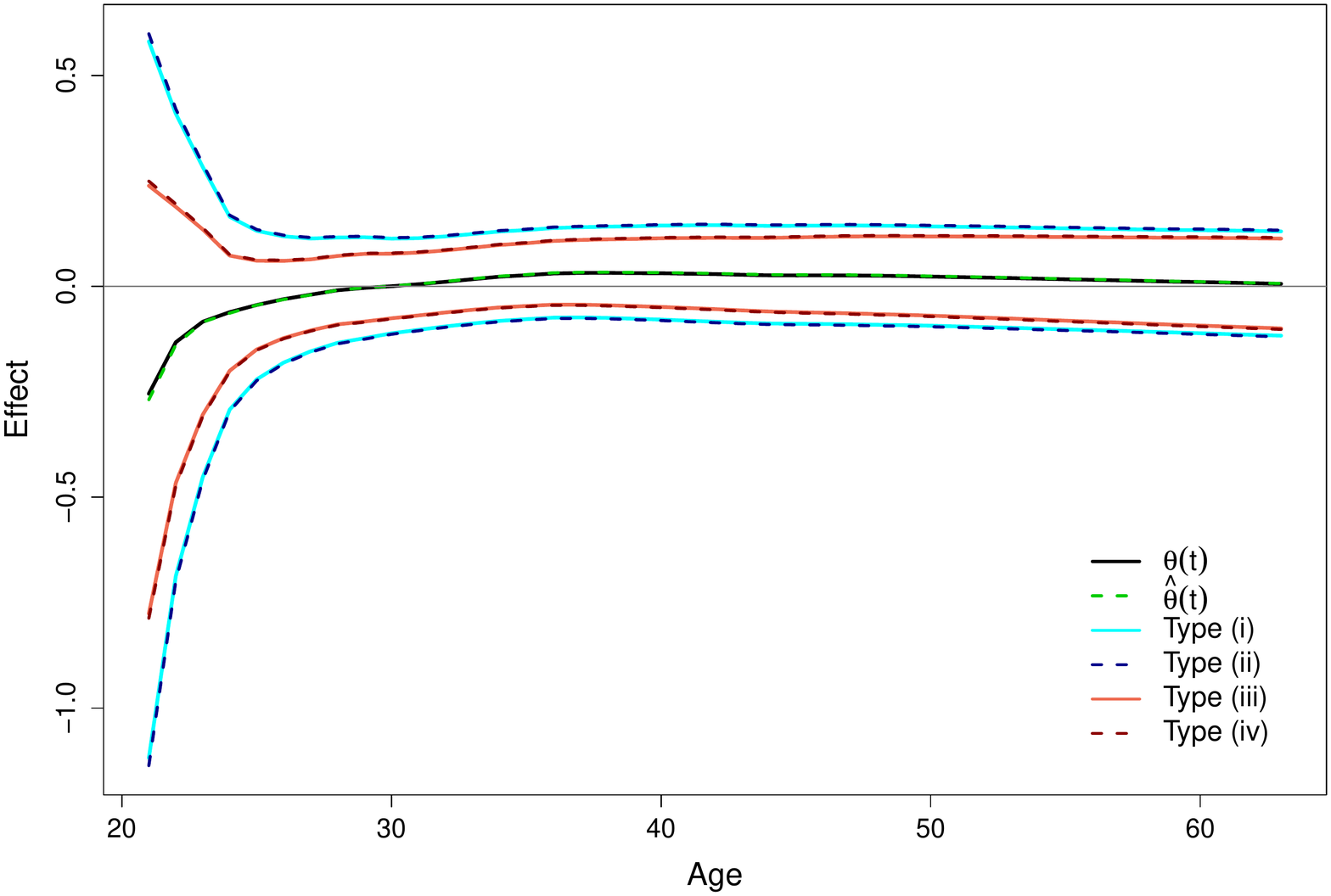} 
\caption{Means of the FATE estimates and four types of confidence bands over the 1000 simulation replicates for male subset, n = 500.}
\label{fig:SimMeanCurves}
\end{figure}

Figure \ref{fig:SimMeanCurves} shows the averages of the estimated FATE and the four different types of confidence bands, taken over 10000 replicates of the simulation study in section \ref{SIM}. The results are for the male subset and a sample size of $n = 500$ in the fixed design setup, corresponding to the second column in Table 1.

\end{appendix}

\end{document}